\documentclass[prb,showpacs,preprintnumbers,amsfonts,amssymb,amsmath,floats,twocolumn,aps,superscriptaddress]{revtex4}

\usepackage{graphicx}
\usepackage{dcolumn}
\usepackage{bm}
\usepackage[final,dvips]{epsfig}
\usepackage{color}
\usepackage{amsmath,amssymb}

\newcommand{\ff}[1]{{\bm #1}}
\newcommand{\Tr}{\mbox{Tr}}
\newcommand{\fig}[1]{Fig.\,\ref{fig:#1}}

\begin{document} 
\sloppy

\title{Phase separation and competing superconductivity and magnetism in the
two-dimensional Hubbard model: From strong to weak coupling}

\author{M. Aichhorn} 
\email{aichhorn@physik.uni-wuerzburg.de}
\affiliation{
Institute for Theoretical Physics and Astrophysics, University of W\"urzburg, Am Hubland, 97074~W\"urzburg, Germany
}

\author{E. Arrigoni}
\affiliation{
Institute of Theoretical Physics and Computational Physics, TU Graz, Petersgasse 16, 8010 Graz, Austria
}

\author{M. Potthoff}
\affiliation{
I. Institute for Theoretical Physics, University of Hamburg, Jungiusstraße 9, 20355 Hamburg, Germany
}

\author{W. Hanke}

\affiliation{
Institute for Theoretical Physics and Astrophysics, University of W\"urzburg, Am Hubland, 97074~W\"urzburg, Germany
}
\affiliation{
Kavli Institute for Theoretical Physics, University of California, Santa Barbara, CA 93106, USA
}

\date{\today}

\begin{abstract}
Cooperation and competition between the antiferromagnetic, d-wave superconducting
and Mott-insulating states are explored for the two-dimensional Hubbard model 
including nearest and next-nearest-neighbor hoppings at zero temperature.
Using the variational cluster approach with clusters of different shapes and sizes 
up to 10 sites,
it is found that the doping-driven transition from a phase with microscopic coexistence of
antiferromagnetism and superconductivity to a purely superconducting phase is discontinuous
for strong interaction and accompanied by phase separation. 
At half-filling the system is in an antiferromagnetic Mott-insulating state with vanishing
charge compressibility. 
Upon decreasing the interaction strength $U$ below a certain critical value of roughly $U_c \sim 4$ 
(in units of the nearest-neighbor hopping), however, the filling-dependent magnetic transition changes 
its character and becomes continuous. 
Phase separation or, more carefully, the tendency towards the formation of inhomogeneous
states disappears. 
This critical value is in contrast to previous studies, where a much larger value was obtained.
Moreover, we find that
the system at half-filling undergoes the Mott transition from an insulator
to a state with a finite charge compressibility 
{\em at essentially the same value}.
The weakly correlated state at half-filling exhibits superconductivity microscopically
admixed to the antiferromagnetic order. 
This scenario suggests a close relation between phase separation and the Mott-insulator physics.
\end{abstract}
\pacs{71.10.-w,71.10.Hf,74.20.-z}

\maketitle

\section{Introduction}

One of the standard models in theoretical studies of high-temperature superconductors 
is the single-band Hubbard model. 
Substantial progress in the understanding of its ground-state properties in two dimensions
has been achieved by applying dynamical quantum-cluster approaches \cite{ma.ja.05.rmp}, such 
as the dynamical cluster approximation~\cite{he.ta.98}, the cellular dynamical mean-field
theory~\cite{ko.sa.01,li.ka.00} or the variational cluster approach (VCA)~\cite{po.ai.03}.
Several studies predict $d$-wave superconductivity at low~\cite{ma.ja.05} or zero 
temperatures~\cite{se.la.05,ai.ar.05,ai.ar.06,ai.ar.06.2,ca.ko.06} 
for intermediate interaction strengths of the order of the free bandwidth.

Experimentally, it is well known that cuprate-based high-$T_C$ compounds at 
low doping concentrations tend to form charge and spin inhomogeneities, such as
stripe~\cite{tr.st.95,tr.ax.96} or checkerboard 
modulations~\cite{ve.mi.04,ho.hu.02,ha.lu.04,mce.le.05}. 
An unbiased theoretical study of inhomogeneous phases is hardly possible
within quantum-cluster approches for the presently accessible cluster sizes.
Furthermore, for a realistic modelling of charge inhomogeneities additional 
non-local interaction terms should actually be taken into account \cite{lo.em.94,ar.ha.02,ki.bi.03}.
However, by searching for phase separation, an overall tendency towards the 
formation of inhomogenous phases can easily and reliably be detected even if only 
homogeneous solutions are allowed within the cluster mean-field calculation. 

Phase separation (PS), i.\,e., separation into two homogeneous {\em macroscopic} regions 
with different thermodynamical properties, is the most simple kind of inhomogeneity.
It has been argued~\cite{lo.em.94,ki.bi.03} that phase separation occurring 
in Hubbard- or $t$-$J$-type models with short-range interactions only, can transform 
to {\em microscopically} inhomogeneous structures, such as stripes,
when long-range repulsions were included~\cite{ar.ha.02}. 
For this reason, the investigation of PS in the Hubbard model is a reasonable 
starting point for an understanding of inhomogeneous phases in the high-$T_{\rm C}$ 
cuprate materials.

The occurrence of phase separation in the Hubbard model has been investigated in the 
past using several different techniques. Calculations for finite clusters using  
quantum Monte-Carlo techniques~\cite{mo.sc.91,be.ca.00} found no evidence for PS
in the Hubbard model with nearest-neighbor hopping only. 
Different results have been obtained in infinite dimensions within dynamical mean-field
theory (DMFT) where PS has been 
reported~\cite{zi.pr.02,we.mi.07,ec.ko.07}. 
PS has also been discussed in the context of {\em marginal quantum criticality}\cite{imad.05,mi.im.07}, 
where PS occurs at the first-order side of the marginal quantum critical point (MQCP).
For the two-dimensional model including hopping between next-nearest neighbors, 
calculations within the dynamical cluster approximation
yield phase separation in the paramagnetic state~\cite{ma.ja.06}. 
VCA and cellular DMFT studies predict phase separation between a phase
with long-range antiferromagnetic (AF)
order at low doping and a superconducting (SC) state at high 
doping~\cite{ai.ar.05,ai.ar.06,ai.ar.06.2,ca.ko.06}. 

The new aspect of our work
is to systematically investigate the fate of the 
phase-separated state within VCA when decreasing the interaction strength. 
We demonstrate by using as reference systems a
variety of cluster sizes up to
10 sites, that the tendency towards PS is lost at small $U$ of the order of
$U_c\sim 4$. Above of this value we found strong evidence from this systematic
study that the inhomogeneous state 
is indeed present in the thermodynamic limit.

In previous cellular DMFT work \cite{ca.ko.06}, a transition which is in certain aspects 
similar to ours was found 
around $U_c\approx 8$, using $2\times 2$ clusters only. 
According to previous studies \cite{pr.ha.95,gr.za.98u,se.tr.04} 
it is known, however, that a physical transition
for the 2D Hubbard model is taking place at much smaller values, $U\approx 4$. Here, the
two separate energy scales $U$ and $J$ are eventually merging, as examplified by the transition 
from two energy bands (coherent low-lying and incoherent Hubbard bands) to just one
single band. A motivation for our study was partly to investigate the relation between the 
metal-to-insulator transition and PS. 
Most strikingly, we find that the transition to a metallic state
at half-filling occurs {\em at 
essentially the same value of $U$}, below which PS disappears. 
This strongly suggests a definite
relation between PS and the Mott insulator, a relation that was previously
speculated about \cite{le.ki.03}.

The paper is organized as follows: In Sec.~\ref{sec:meth}, we introduce the 
model and briefly review the variational cluster approximation.
In Sec.~\ref{sec:res}, our results are presented and discussed.
The conclusions are summarized in Sec.~\ref{sec:concl}.

\section{Theory}\label{sec:meth}

Using standard notations, the Hubbard model is given by $H=H_{0}(\ff t)+H_{1}$ where
\begin{align}
  H_{0}(\ff t)&=-t_{n.n.}\sum_{\langle ij\rangle,\sigma}c_{i\sigma}^\dagger c_{j\sigma}^{\phantom{\dagger}} - 
  t_{n.n.n.} \sum_{\langle\!\langle ij\rangle\!\rangle,\sigma}c_{i\sigma}^\dagger c_{j\sigma}^{\phantom{\dagger}}\\
  H_{1} &= U\sum_in_{i\uparrow}n_{i\downarrow}.
\end{align}
The operator $c_{i\sigma}^{(\dagger)}$ creates (annihilates) an electron with spin $\sigma$ 
at the site $i$, and
$n_{i\sigma}$ is the corresponding occupation number operator. 
We consider both hopping $t_{n.n.}$ along nearest-neighbor bonds 
$\langle ij\rangle$ as well as hopping $t_{n.n.n.}$ along next-nearest-neighbor bonds 
$\langle\!\langle ij\rangle\!\rangle$.
$U$ is the local Coulomb interaction. We set the unit of energy by $t_{n.n.}$ and choose $t_{n.n.n.}=-0.3t_{n.n.}$ 
throughout the paper which is a realistic value for the cuprate materials. 

The main idea of the variational-cluster approximation (VCA)~\cite{po.ai.03,pott.03.epjb2} is to consider
a ``reference system'' to span a space of trial self-energies among which the self-energy that
describes best the physics of the infinite-size lattice model is obtained via a dynamical 
variational principle $\delta \Omega[\ff \Sigma]=0$. Here $\Omega$ stands for 
the grand potential.
The reference system is given by a Hamiltonian $H'$ with the same interaction part $H_{1}$
as the physical system but with modified
one-particle parameters $\ff t'$, i.\,e., $H'=H_{0} (\ff t') + H_{1}$. 
Within the VCA one takes as a reference system a lattice split up into
isolated clusters of a given size.
Thereby, the effects of short-range correlations on the self-energy are included on a
scale given by the cluster extension.
Trial self-energies $\ff \Sigma = \ff \Sigma(\ff t')$ are varied by varying the
parameters $\ff t'$. 
Inserting the trial self-energy into the self-energy functional generates a function
$\Omega(\ff t') = \Omega[\ff \Sigma(\ff t')]$ the stationary points of which we are
interested in.
It can be shown \cite{po.ai.03,pott.03.epjb2} that
\begin{equation}\label{eq:ocalc} 
        \Omega (\ff t^\prime) 
        =\Omega^\prime + \Tr \ln (\ff G_{0,\ff t}^{-1}-\ff \Sigma(\ff t^\prime))^{-1}
        - \Tr \ln \ff G_{\ff t^\prime} \: ,
\end{equation}
where $\Omega^\prime$ is the grand potential and $\ff G_{\ff t^\prime}$ the Green's 
function of the reference system and $\ff G_{0,\ff t}$ the non-interacting Green's function 
of the physical system.
While $\ff G_{0,\ff t}$ is easily accessible,
we calculate the reference system's properties using full diagonalization for small clusters, 
and the band Lanczos methods for larger ones. 
A description of the numerical details can be found in Ref.~\onlinecite{ai.ar.06.2}.

Since we are interested in PS involving symmetry-broken antiferromagnetic and 
$d$-wave superconducting phases, our reference system includes the corresponding 
(ficticious) symmetry-breaking fields,
\begin{subequations}
  \begin{align}\label{eq:varpar}
    H'_{\rm AF} &= h'_{\rm AF} \sum_{i\sigma} (n_{i\uparrow} - n_{i\downarrow}) 
    e^{i \ff{Q} \ff{R_i}} \\
    H'_{\rm SC} &= h'_{\rm SC} \sum_{ij} \frac{\eta_{ij}}{2} 
    (c_{i\uparrow} c_{j\downarrow} + \mbox{h.c.})
  \end{align}  
\end{subequations}
where $h'_{\rm AF}$ and $h'_{\rm SC}$ are the strengths of a staggered magnetic
and of a nearest-neighbor $d$-wave pairing ``Weiss'' field, respectiely.
Furthermore, $\ff{Q} = (\pi, \pi)$ is the AF wave vector, and $\eta_{ij}$ 
denotes the $d$-wave form factor which is 
equal to $+1$ ($-1$) for nearest-neighbor sites with $\ff R_i - \ff R_j$ in $x$ ($y$) direction. 
In addition, an on-site potential is included in the set of variational parameters to ensure
a thermodynamically consistent determination of the average particle number~\cite{ai.ar.06}.
Note that all three variational parameters couple to one-particle operators only.

\section{Results}\label{sec:res}

\begin{figure}[t]
  \centering
  \includegraphics[width=0.7\columnwidth]{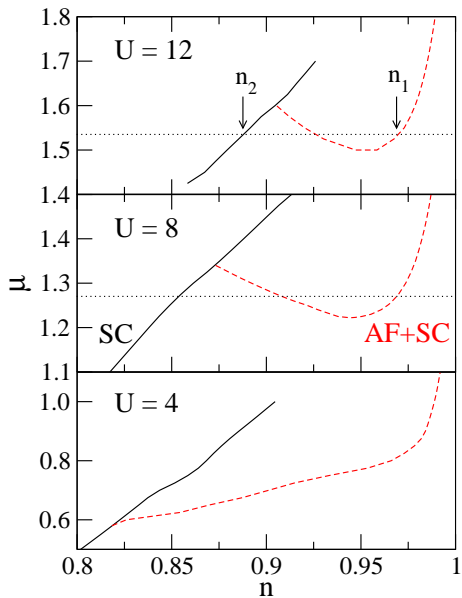}
  \caption{\label{fig:muvsn}%
    (Color online) 
    Chemical potential $\mu$ as function of the average density $n$, calculated with a $2\times 2$ cluster as reference system
    for different values of $U$. From top to bottom: $U=12$, 8, and 4, respectively. 
    $n_1$ and $n_2$ denote the boundaries of the instability (PS) region.
    Black solid lines: Pure SC phase. Red dashed
    lines: AF+SC mixed phase. Dotted horizontal line: 
    critical chemical potential (where applicable).
  }
\end{figure}

For the calculations we concentrate on the hole-doped side of the phase diagram since it has been shown 
that there PS is much more pronounced than in the electron-doped case~\cite{ai.ar.05,ai.ar.06,ai.ar.06.2}.
The occurence of PS can 
be best inferred from the dependence of 
the chemical potential on the particle number.
For a physical system in thermodynamical equilibrium the charge susceptibility 
$\kappa = \partial n / \partial \mu$ must be non-negative. 
Hence, $\kappa<0$ indicates a thermodynamically unstable phase.
At a fixed {\em average} density $n$ lying in the instability region $n_1 < n < n_2$,
the free energy can be reduced if the system develops two spatially separated homogeneous
phases, one with a fraction $x=\frac{n_2-n}{n_2-n_1}$ of particles at the density $n_1<n$ and 
another one with the fraction $1-x$ at $n_2>n$, rather than having a single homogeneous phase.
The boundaries $n_1$ and $n_2$ of the instability region can be obtained by a Maxwell
construction (see Fig.~\ref{fig:muvsn}), in close analogy to a gas-liquid system.
After the Maxwell construction, the {\em physical} chemical potential
is independent of $n$ between $n_1$ and $n_2$.

In \fig{muvsn} we show results for the interrelation of the chemical potential $\mu$ 
and the average density $n$ as a function of Coulomb interaction $U$.
The calculations have been done with a
$2\times 2$ cluster as a reference system. In the strong coupling regime, $U=12$, we find
that the chemical potential $\mu$ as function of the density $n$ shows 
a nonmonotonic behavior (and thus $\kappa<0$) as mentioned above.
We conclude that for this coupling and within the precision set by the $2\times 2$ 
cluster reference system, the results imply PS into an AF+SC mixed phase~\cite{afsc} at
low doping and a purely SC phase at higher doping. The critical chemical
potential where the two phases coexist, is indicated by a horizontal dotted line
in \fig{muvsn}. This behavior is very similar to the previously reported one for
$U=8$~\cite{ai.ar.06,ai.ar.06.2}. 
For comparison, results for $U=8$ -- as published in Ref.~\onlinecite{ai.ar.06} -- are shown 
in \fig{muvsn} in the middle panel. 

The picture changes even qualitatively when going to weak interactions, e.\,g., $U=4$. As shown
in the lower panel of \fig{muvsn}, the chemical potential as function of $n$ shows 
a positive slope {\em for all dopings}, without an intermediate region with negative sign.
We conclude that for weak interaction there is no tendency to the formation of inhomogeneities, 
even for the smallest cluster we used for our calculations. Instead, the AF+SC solution is stable up
to larger dopings, and the staggered magnetization vanishes continuously in the {\em stable}
and {\em homogeneous} solution.

\begin{figure}[t]
  \centering
  \includegraphics[width=0.5\columnwidth]{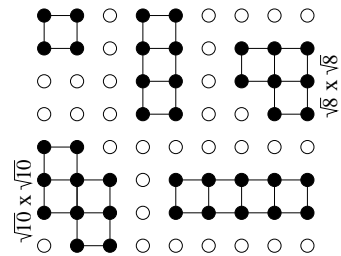}
  \caption{\label{fig:clusters}%
  Clusters used as reference systems in this study. The 
  ``$\sqrt{10}\times\sqrt{10}$'' and ``$\sqrt{8}\times\sqrt{8}$''
  clusters are marked explicitly in the figure.}
\end{figure}

Since the occurrence of PS is subject to rather strong finite-size effects~\cite{ai.ar.06.2}, and
in order to further elucidate the difference between the coupling regimes, 
we have recalculated the phase diagram using larger clusters with $L_c=8$ and $L_c=10$ lattice 
sites as reference systems (see \fig{clusters}). For both $L_c=8$ and $10$ we have considered
two different cluster geometries to estimate dependencies on the cluster shape.

\begin{figure}[t]
  \centering
  \includegraphics[width=0.8\columnwidth]{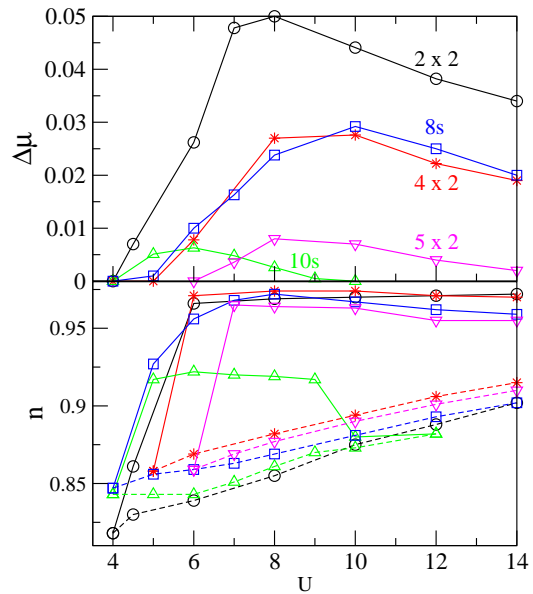}
  \caption{\label{fig:critical}%
    (Color online) Characteristic energy
    $\Delta\mu=\mu^\ast - \mu_c$ (top) and critical densities $n_1$ and $n_2$ (bottom) as functions
    of $U$, calculated using the clusters shown in \fig{clusters} 
    as reference systems, $2\times 2$ (black, circles), $\sqrt{8}\times\sqrt{8}$ (blue, squares),
    $4\times 2$ (red, stars), $5\times 2$ (magenta, triangles down), and $\sqrt{10}\times\sqrt{10}$ 
    (green, triangles up). 
    Lower plot: Solid lines correspond to the critical density $n_1$,
    dashed lines to the critical density $n_2$.
  }
\end{figure}

\begin{figure}[t]
  \centering
  \includegraphics[width=0.7\columnwidth]{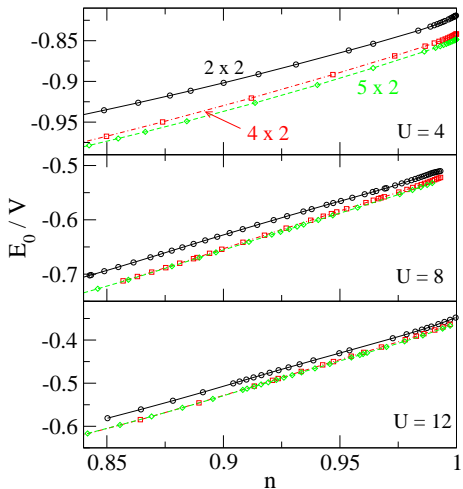}
  \caption{\label{fig:e0}%
    (Color online) Ground-state energy per site $E_0/L$ as function of electron density $n$, for $U=4$ (top), $U=8$ (middle),
    and $U=12$ (bottom). Results are shown for the $2\times 2$ (black solid), $4\times 2$ 
    (red dashed-dotted), and $5\times 2$ (green dashed) reference systems.}
\end{figure}

In addition to the boundaries of the instability region, $n_1$ and $n_2$, we define a characteristic 
energy $\Delta \mu = \mu^\ast - \mu_c$. 
Here, $\mu^\ast$ is defined as the point where the slope of $\mu(n)$ changes sign, and $\mu_c$ is the 
critical chemical potential, i.\,e., the point at which $\Omega_{\rm AF+SC}(\mu)$ crosses $\Omega_{\rm SC}(\mu)$.
Equivalently, $\mu_c$ is fixed by the Maxwell construction, see \fig{muvsn}.
In the limit of infinite cluster size, the characteristic energy $\Delta \mu$ must vanish as the 
reference system (and thereby also the original system) is solved exactly within the VCA: 
For $L_c\to\infty$ (and for densities in the exact instability region) the reference system 
{\em spontaneouly} generates the phase-separated state, and $\mu(n)$ becomes flat between 
$n_1$ and $n_2$, as discussed above.
On the other hand, the difference between the critical densities $n_2-n_1$ must converge to 
a nonzero value for $L_c\to\infty$ whenever the system in the thermodynamic limit shows
phase separation.
Notice that the difference between the grand potentials $\Omega_{\rm AF+SC}$ and
$\Omega_{\rm SC}$ becomes 
smaller and smaller with increasing cluster size. Thus, together with
the increasing numerical effort, this makes the identification of $\Delta \mu$ harder for larger
clusters.
Contrary, the slope of $\Omega(\mu)$, i.\,e. the 
particle density, is not affected by a systematic shift and,
therefore, its calculation is quite reliable also for larger $L_c$.

Results for these quantities are shown in \fig{critical}. The upper panel shows  
$\Delta \mu$ as a function of the Hubbard interaction $U$. As expected from the
discussion above, $\Delta\mu$ decreases for increasing cluster size.
Quite surprising, however, is the dependence of $\Delta \mu$ on the interaction:

As $\Delta \mu \neq 0$ is a finite-size effect, one could expect $\Delta\mu$ to be
smaller for stronger interactions since a cluster mean-field approach quite generally
is expected to be more reliable in the strong-coupling than in the weak-coupling
regime.
At least, it is 
known from previous calculations~\cite{ai.ev.04}
that for weaker interactions finite-size effects are larger as could be seen
from the stronger dependence of the optimal variational parameters on the size 
of the reference system cluster.
In fact, we find for large $U$ (see \fig{critical}) that $\Delta\mu$ increases with
decreasing interaction. 
Below $U \approx 10$, however, $\Delta\mu$ starts to decrease and eventually even 
vanishes at some critical value $U_c$ (which exhibits a finite but weak dependence 
on the cluster size).
This decrease indicates a qualitative change of the phase transition with decreasing $U$.

From the analysis of the critical densities we 
see that the vanishing of 
$\Delta \mu$ corresponds to the disappearance of the phase-separated state:
The critical densities $n_1$ and $n_2$ of the two coexisting phases are shown in 
the lower panel of \fig{critical}.
The dependence of their difference $n_2-n_1$ 
on $U$ closely resembles the dependence of $\Delta\mu$.
In particular, for $U$ below some $U_c$, $n_1$ and $n_2$ collapse to a
single point, i.\,e., phase separation disappears.

\begin{figure}[t]
  \centering
  \includegraphics[width=0.95\columnwidth]{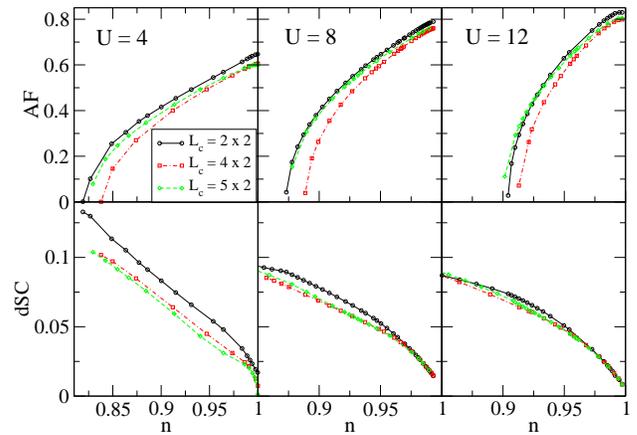}
  \caption{\label{fig:orderpar}%
    (Color online) Order parameters as a function of density 
    for $U=4$ (left), $U=8$ (middle), and $U=12$ (right). Top: AF order parameter. 
    Bottom: d-wave SC order parameter. Results are shown for the $2\times 2$ (black solid), $4\times 2$ 
    (red dashed-dotted), and 
    $5\times 2$ (green dashed) reference systems.
  }
\end{figure}

It is encouraging to see that, except for the $\sqrt{10}\times\sqrt{10}$ cluster, 
the critical densities shown in this plot depend only very weakly on the cluster size.
The $\sqrt{10}\times\sqrt{10}$ cluster seems to behave differently.
Probably, its shape (see Fig.~\ref{fig:clusters}) makes this cluster unsuitable for 
finite-size scaling.
This is also confirmed by the fact that $\Delta \mu$ does not display the
correct scaling behavior for this cluster. 
In addition, the $5 \times 2$ cluster with the same number of sites $L_c=10$ does 
display the correct $L_c$ dependence of $\Delta \mu$.
Excluding the results from the 
$\sqrt{10}\times\sqrt{10}$ cluster,
$n_1$ and $n_2$ show a very weak dependence on $L_c$.
Close to $U_c$, the $L_c$ dependence is somewhat stronger again which
is not surprising, of course.

In view of the results for different interaction strengths and cluster sizes
we conclude that PS persists in the thermodynamic limit down to a critical 
interaction $U_c$.
Admittedly, there is a rather large uncertainty in the determination of $U_c$: 
Estimates 
range between $U_c \sim 4$ and $U_c\sim 6$.
Interestingly, results from the $\sqrt{10}\times\sqrt{10}$ 
cluster do not only show a lower critical interaction but
also an upper critical value $U_{c2}\approx 10$ above which PS
disappears. 
As discussed above, however, the question is whether results for this cluster
shape are reliable or not.

From the grand potential we can extract the ground-state energy of the the system by
$E_0 = \Omega + \mu N$. Results for the ground-state energy per site, i.\,e. $E_0 / L$, 
are shown in \fig{e0} for $U=4$, 8, and 12, resp.
Although some (small) finite-size
effects are visible, the ground-state energy seems to be well converged in our approach.
As a check for 
consistency we found that the general relation $\partial (E_0 /L) / \partial n =\mu$ is 
fulfilled within numerical accuracy. This relation is a consequence of the Legendre transformation
$E_0=\Omega + \mu N$ and must be true in all calculations independent of the cluster used
as a reference system.

The filling dependence of the order parameters is displayed in \fig{orderpar}.
Although some finite-size effects are visible, the results strongly indicate that
superconductivity persists in the thermodynamic limit.
We can also extract an interesting trend when 
comparing the 
results for $U=8$ with those for $U=4$:
While the staggered magnetization {\em decreases} the SC order parameter 
{\em increases} with decreasing $U$.
At the same time the doping region where a magnetic solution exists, extends to 
somewhat larger dopings for smaller $U$. 
Another difference is the small but finite value of the dSC order parameter at
half-filling ($n=1$) for $U=4$, as can be seen in the lower right plot of 
\fig{orderpar}. 
This is directly related to the closing of the Mott-Hubbard gap, as discussed below.

There are also some differences between $U=8$ and $U=12$, although they are considerably smaller.
Taking for instance the boundary of the AF phase, i.\,e. the density where the
AF order parameter vanishes, it is at $n\approx 0.82$ for $U=4$, 
$n\approx 0.87$ for $U=8$, and $n\approx 0.89$ for $U=12$. Also the changes of the 
SC order parameter are very small. We want to stress again that the major difference, 
however, is the {\em absence} of
PS for $U=4$, see Figs.~\ref{fig:muvsn} and \ref{fig:e0}.

\begin{figure}[t]
  \centering
  \includegraphics[width=0.7\columnwidth]{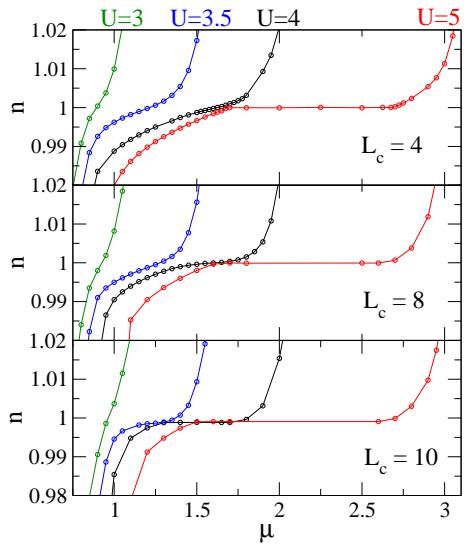}
  \caption{\label{fig:nvsmu}%
    (Color online) Closing of the Mott gap as function of $U$. 
    From top to bottom: $L_c=2\times 2$, $L_c=4\times 2$, and $L_c=\sqrt{10}\times\sqrt{10}$ 
    site clusters. From left to right: $U=3$ (green), 3.5 (blue), 4 (black), and 5 (red). 
  }
\end{figure}

To understand the occurrence of a critical interaction for PS
as well as the finite dSC order parameter at half-filling, we study the $n(\mu)$ behavior 
close to half-filling as a function of $U$. 
Results are shown in \fig{nvsmu} in the vicinity of $n=1$.
For $U=5$ there is a flat $n(\mu)$ dependence for a finite range of chemical potentials, i.\,e.,
a vanishing charge susceptibility $\kappa=0$.
This indicates the presence of an (AF) Mott insulating state at half-filling.
The Mott gap shrinks with decreasing $U$, and eventually the 
system shows metallic behavior at $n=1$. 
This agrees well with previous results~\cite{gr.za.98u,se.tr.04,on.im.03,mi.im.06}.
For the smallest system the transition to the metal takes place at an interaction strength somewhat 
above $U=4$, whereas for larger clusters the critical $U$ is shifted to slightly smaller values.
For the $5\times2$ cluster (not shown) it is slightly below $U=4$, and for the $\sqrt{10}\times\sqrt{10}$ 
cluster the transition takes place slightly above $U=3.5$. 
Notice that the value of $U$ where the gap closes can also be read off 
directly from the dSC order parameter. Below the critical $U$, 
the system is metallic even at $n=1$, and hence can also be superconducting at half-filling.
This is seen in our calculations which give a finite order parameter at
$n=1$ (see \fig{orderpar}). We like to stress that close to half-filling the solution 
with lowest energy is always a mixed AF+SC one~\cite{afsc}. 
We checked this by comparing this solution to the pure AF and SC ones.

It is obvious that our determination of $U_c$ is not as precise as it can be done by
other methods, e.\,g. the path-integral renormalisation group\cite{mi.im.06}. 
Moreover, with the cluster sizes available in the present form of the VCA, it is not possible to 
detect whether or not we have a MQCP\cite{imad.05,mi.im.07}
with a diverging charge susceptibility $\kappa$ at the metal-insulator boundary.
Nevertheless, our
calculation allows to
make a qualitative connection of the MIT and PS. It indicates
that the collapse of the upper and lower dopings $n_1$ 
and $n_2$ (i.\,e. the disappearance of PS) and the closing of the Mott gap 
occur at almost the same critical interaction strength.
It appears that it is important to have a Mott insulator at half-filling 
in order to get phase separation away from half-filling.
Due to the limited cluster sizes available and the according limited accurary in the determination of
critical points, however, this statement is somewhat speculative.
Note that a relation between phase separation and the doping of a Mott insulator
has been discussed in Ref.~\onlinecite{le.ki.03}. 
There, it has been argued that a Mott insulator with broken symmetry, which is the
$SU(2)$ symmetry in our case, enhances the possibility of PS.

\begin{figure}[t]
  \centering
  \includegraphics[width=0.5\columnwidth]{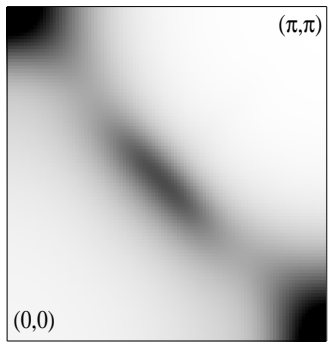}
  \caption{\label{fig:fs}%
    Representation of the Fermi surface for an interaction $U=3.5 < U_c$ and $n=1$. 
    Calculation using the $\sqrt{8}\times\sqrt{8}$ cluster. 
    Dark regions denote large spectral weight integrated over a small frequency 
    window of $\Delta\omega=\pm0.05$ around $\omega=0$. 
  }
\end{figure}

\begin{figure}[t]
  \centering
  \includegraphics[width=0.7\columnwidth]{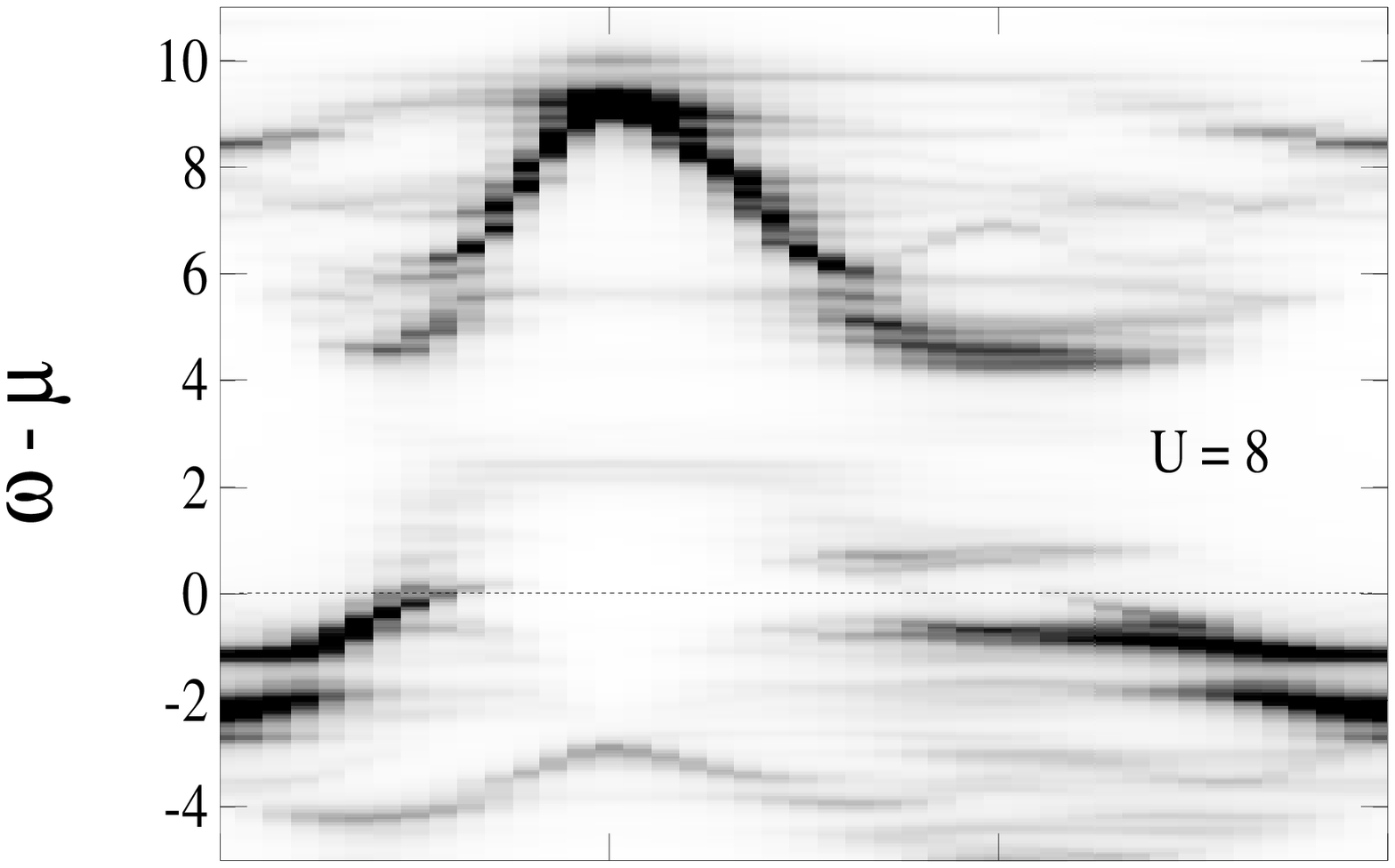}\\
  \includegraphics[width=0.7\columnwidth]{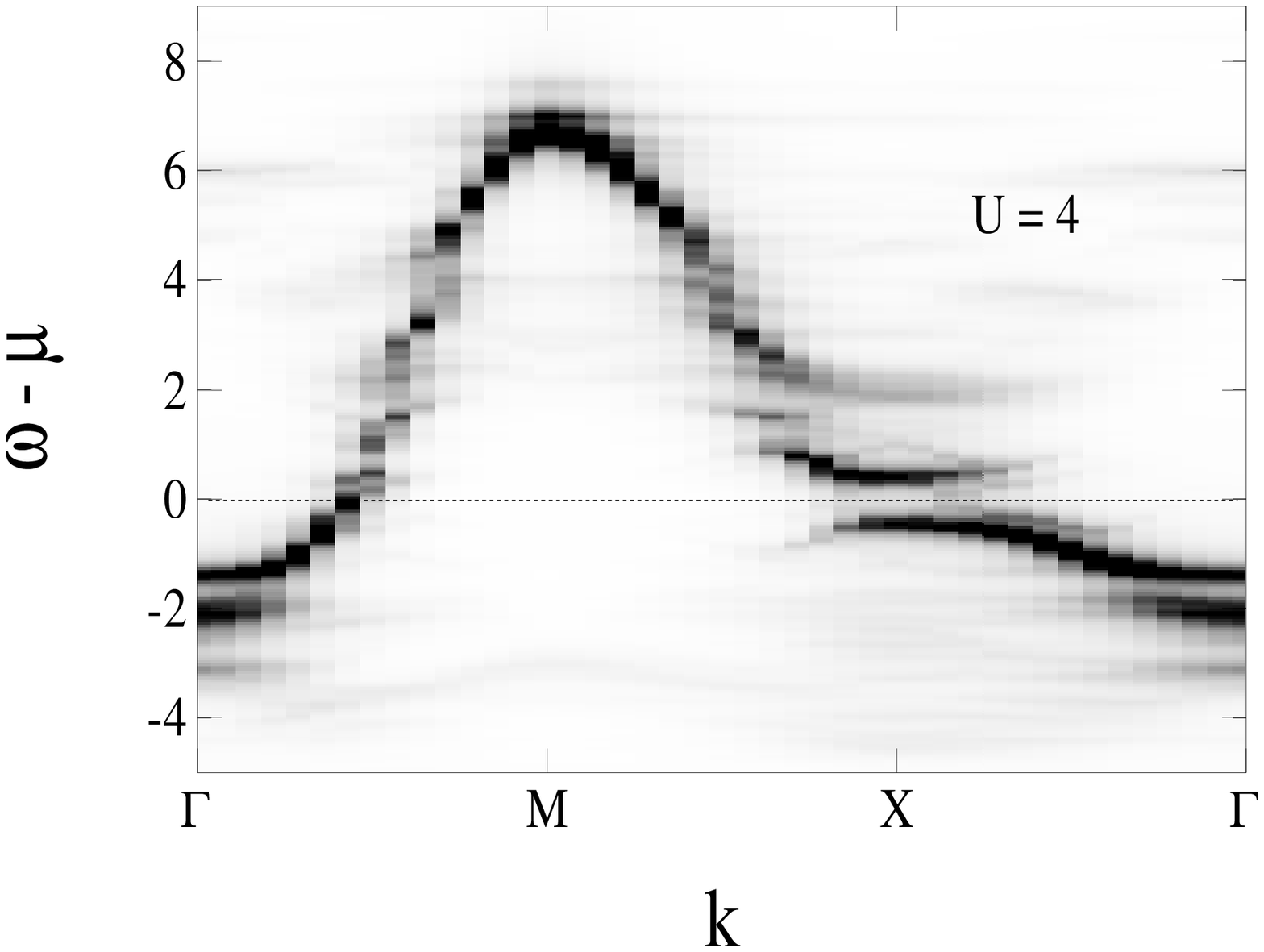}
  \caption{\label{fig:akw}%
    Spectral function for $U=8$ (top) and $U=4$ (bottom). Calculation using a 
    $\sqrt{8}\times\sqrt{8}$ cluster. The horizontal dashed line marks the chemical
    potential.
  }
\end{figure}

In order to analyze the relation with single-particle excitations, we plot in \fig{fs}
the Fermi surface for weak coupling, $U=3.5$, calculated with the $\sqrt{8}\times\sqrt{8}$ cluster.
The closing of the Mott gap is clearly manifested by the occurrence of both hole and electron pockets 
near $(\frac{\pi}{2},\frac{\pi}{2})$ and $(\pi,0)$, respectively, in direct
contrast to the Fermi surface for 
larger coupling and $U=8$ where one has {\em either} hole {\em or} electron pockets at low
doping~\cite{ai.ar.06}. The simultaneous occurrence of hole and electron pockets has also been reported
recently in weak-coupling calculations close to half-filling~\cite{re.ro.07}. 

Another way to see the closing of the Mott gap is to look at the spectral function directly, which we plot
in \fig{akw} for $U=8$ and $U=4$, resp. Calculations have been done using the same $\sqrt{8}\times\sqrt{8}$ cluster
as for \fig{fs}, but at finite hole-doping just at the critical doping $n_2$ in the pure SC phase. From this
figure, it is obvious that for $U=8$ one has a quasi-particle band at the Fermi level, well separated from the
upper Hubbard band. For $U=4$, however, the upper and lower Hubbard bands merge, and closely resemble the
band structure of the free system, except for the superconducting gap around $(\pi,0)$.

Related ideas of a change of the type of the phase transition as function of the interaction strength
have been reported quite recently based on cellular DMFT calculations
by Capone et al.~\cite{ca.ko.06} for the two-dimensional Hubbard model with $t_{n.n.n.}=0$. 
Capone et al.\ inferred a
critical interaction of approximately $U_c\approx 8$ 
below which the 
doping-dependent transition to a paramagnet is continuous and above which there is a
first-order transition accompanied by phase separation.
The $U_c$ reported for $t_{n.n.n.}=0$ and using cellular DMFT with $2\times 2$ clusters is somewhat 
larger as in our study.
A qualitative difference, however, is that
we always find a mixed AF+SC state at low doping whereas the authors of Ref.~\onlinecite{ca.ko.06} 
find a pure AF phase for strong interaction close to half-filling. 
One should note, however, that finding the stable solution with lowest energy can be a 
difficult task within quantum-cluster approaches.
In Ref.~\onlinecite{ca.ko.06} the calculation starts from pure AF and SC
solutions and a mixed AF+SC solution is searched for by applying small perturbations 
to the pure ones. 
This procedure does not necessarily lead from one local solution to another, 
especially if the two are well separated in parameter space.

Finally, we would like to comment on the relation of our results to the $t$-$J$-model. Translating
our results of \fig{critical} to $J\propto \frac{t^2}{U}$, we argue that PS for small
$J$ (large $U$) should be weak, or eventually absent in the thermodynamic and $U\to \infty$ limit, 
since $\Delta \mu$ decreases significantly with increasing $U$. However, from our results it is not 
possible to deduce a definite critical $J_c$ below which PS should be absent. Nevertheless, the
spectral function and in particular the low-lying quasi-particle band in our calculations is in well 
agreement with the low-lying bands in the $t$-$J$-model.\cite{ze.pr.07u}

\section{Conclusions}\label{sec:concl}

To summarize, we have studied the occurrence of phase separation in the two-dimensional
Hubbard model at zero temperature depending on the strength of the Hubbard 
interaction $U$. 
We have employed the variational-cluster approach using clusters of different shapes
and sizes up to 10 sites. 
Our results show that the nature of the 
doping-dependent transition from 
the antiferromagnetic and superconducting state to a non-magnetic and purely 
superconducting state changes from discontinuous to continuous when going from 
the strong- and intermediate- to the weak-interaction regime.
Below a critical value for the interaction strength
no phase separation can be found. 
Above the critcal value, there is a clear discontinuity in the density as a function
of the chemical potential with a small finite-size error only.
This is a strong indication that 
PS, or at least a phase with mesoscopic inhomogeneities, 
should persist in the thermodynamic limit, as long as the interaction is sufficiently strong.

We also studied the interaction-driven Mott transition at half-filling.
For weak interactions (and $t_{n.n.n.}=-0.3t_{n.n.}$) the system is in a mixed AF+dSC state
with a finite charge susceptibility. 
The transition to the AF Mott insulator with vanishing charge susceptibility takes place at 
$U\approx 3.5-4$. 
As the Mott transition at half-filling takes place at the same interation strength where phase 
separation appears in the doped case, one might speculate that these two phenomena are closely 
related.

\begin{acknowledgments}
The authors thank M.~Imada for his valuable comments on the manuscript.
This work has been supported by the Deutsche Forsch\-ungsgemeinschaft within the Forschergruppe
FOR~538, by the Austrian science fund (FWF project P18551-N16), and in part by the National 
Science Foundation under Grant No. PHY05-51164.
\end{acknowledgments}


\begin{thebibliography}{10}
\bibitem{afsc}
A mixed AF+SC phase has been first obtained by cluster methods in
  Ref.~\onlinecite{li.ka.00}, and has been observed experimentally in
  Ref.~\onlinecite{da.ka.05} and recently in Ref.~\onlinecite{yu.hi.07u}.
\end{thebibliography}
\end{document}